# Low temperature photoluminescence study of thin epitaxial GaAs films on Ge substrates.

Guy Brammertz, Yves Mols, Stefan Degroote, Vasyl Motsnyi, Maarten Leys, Gustaaf Borghs, and Matty Caymax.
Interuniversity Microelectronics Center (IMEC vzw), Kapeldreef 75, B-3001 Leuven, Belgium.

**Abstract**

Thin epitaxial GaAs films, with thickness varying from 140 to 1000 nm and different Si doping levels, were grown at 650°C by organometallic vapor phase epitaxy (OMVPE) on Ge substrates and analyzed by low-temperature photoluminescence (PL) spectroscopy. All spectra of thin GaAs on Ge show two different structures, one narrow band-to-band (B2B) structure at an energy of ~1.5 eV and a broad inner-bandgap (IB) structure at an energy of ~1.1 eV. Small strain in the thin GaAs films causes the B2B structure to be separated into a light-hole and a heavy-hole peak. At 2.5 K the good structural quality of the thin GaAs films on Ge can be observed from the narrow excitonic peaks. Peak widths of less than 1 meV are measured. GaAs films with thickness smaller than 200 nm show B2B PL spectra with characteristics of an n-type doping level of approximately $10^{18}$ atoms/cm$^3$. This is caused by heavy Ge diffusion from the substrate into the GaAs at the heterointerface between the two materials. The IB structure observed in all films consists of two gaussian peaks with energies of 1.04 eV and 1.17 eV. These deep trapping states arise from Ge-based complexes formed within the GaAs at the Ge-GaAs heterointerface, due to strong diffusion of Ge atoms into the GaAs. Because of similarities with Si-based complexes, the peak at 1.04 eV was identified to be due to a $Ge_{Ga}$-$Ge_{As}$ complex, whereas the peak at 1.17 eV was attributed to the $Ge_{Ga}$-$V_{Ga}$ complex. The intensity of the IB structure decreases strongly as the GaAs film thickness is increased. PL intensity of undoped GaAs films containing anti phase domains (APDs) is four orders of magnitude lower than for similar films without APDs. This reduction in intensity is due to the electrically active Ga-Ga and As-As bonds at the boundaries between the different APDs. When the Si-doping level is increased, the PL intensity of the APD-containing films is increased again as well. A film containing APDs with a Si doping level of ~$10^{18}$ atoms/cm$^3$ has only a factor 10 reduced intensity. We tentatively explain this observation by Si or Ge clustering at anti phase boundaries, which eliminates the effects of the Ga-Ga and As-As bonds. This assumption is confirmed by the fact that, at 77 K, the ratio between the intensity of the IB peak at 1.17 eV to the intensity of the peak at 1.04 eV is smaller than 1.4 for all films containing APDs, whereas it is larger than 1.4 for all films without APDs. This shows stronger clustering of Si or Ge in the material with APDs. For future electronic applications, Ge diffusion into the GaAs will have to be reduced. PL analysis will be a rapid tool for studying the Ge diffusion into the GaAs thin films.

# 1. Introduction

As Si device scaling for future generations of CMOS circuits becomes increasingly difficult [Ieong 04], an alternative option becomes more and more appealing: GaAs and Ge are intrinsically faster semiconductors than Si [Kittel 96] and especially the integration of Ge PMOS with GaAs NMOS is a very attractive replacement for fast Si CMOS structures. Therefore, thin GaAs epitaxy on Ge will be a key requirement for the fabrication of co-integrated GaAs and Ge MOS devices. GaAs on Ge film thickness smaller than 100 nm should be achieved with very good epitaxial properties. Unfortunately for these small film thickness, a lot of characterization techniques, like X-ray diffraction, reach their limits, whereas other characterization techniques, like Transmission Electron Microscopy, adapted to this film thickness are very costly and time consuming.

Photoluminescence spectroscopy (PL) could be a very rapid and easy method to access the quality of the very thin GaAs layers on Ge substrates. Whereas the method by itself is non-destructive and not very time consuming, it gives a lot of information about the structural quality of the material as well as information about doping materials and concentration, as all these points influence the behaviour of the charge carriers within the material.

In the following, we first describe the experimental low temperature PL setup, followed by a short description of the MOCVD process used to grow the high quality GaAs layers on Ge substrates. Then, some theoretical aspects about the formation of the GaAs/Ge heterostructure, photon absorption and carrier diffusion inside the material are highlighted.

We then describe the band-to-band PL structure of anti phase domain-free layers of GaAs on Ge. First, results on undoped GaAs are discussed, followed by an analysis of doped GaAs films on Ge. As a third point, very thin undoped GaAs on Ge layers are discussed, as these present some unusual behaviour.

Subsequently, the PL spectra emitted from deep trapping states within the bandgap of the GaAs on Ge are presented. Here, the variation of these structures with film thickness and doping concentration are discussed.

Finally, the PL from GaAs on Ge with anti phase domains is described as well as the variation of the latter as the Si-doping of the layers is increased.

# 2. Experimental set-up of PL experiment

An experimental set-up was mounted in the Nanolab on the ground floor of IMEC 1. A schematic of the optical table is shown in Fig.1.

The light source consists of a ~15 mW Ar laser of which the wavelength of 488 nm (2.54 eV) is filtered out. An optical chopper periodically interrupts the laser beam at a speed of approximately 120 Hz. The output signal of the Chopper is used as the reference signal for the EG&G lock-in amplifier. Via a small mirror the beam is directed almost perpendicularly onto the sample, which is mounted inside a cryostat. The illuminated area of the sample is slightly larger than 1 mm$^2$, which corresponds to an illumination density of approximately 1 W/cm$^2$. The cryostat is a Janis Research He-flow cryostat, which can be operated down to 2K. The PL light emitted from the sample is captured by a first 15 cm focal length lens with a diameter of 8 cm. A second 8 cm diameter lens with a focal length of 80 cm focuses the parallel beam of light into the monochromator entrance slit. The monochromator is a 640 mm Jobin-Yvon grating spectrometer. The mounted grating is a 600 grooves/mm grating and is optimised for a wavelength of 750 nm. A personal computer standing next to the set-up controls the position of the grating and accordingly determines the wavelength at the exit slit of the monochromator. A mirror positioned in front of the exit slits can direct the light either onto a Si or a Ge detector, which are read out by the lock-in amplifier. The Si detector is operational for light with a wavelength ranging from 300 to 1100 nm, whereas the Edinburgh Instruments liquid-$N_2$ cooled Ge detector operates in the wavelength range 800 to 1700 nm. This set-up allows us to scan the energy range from 0.75 eV upwards. A custom-made Labview control software allows the automatic acquisition of spectra and read-out of all data. A screenshot of the control software is shown in figure 2.

## 3. GaAs MOCVD on Ge substrates

The films analysed by PL were deposited in the laboratories of the MCP department in the support area of IMEC 1. The deposition tool is a Thomas Swan metal-organic chemical vapour deposition (MOCVD) reactor. The deposition chamber consists of a vertical flow reactor with a 3-zone carbon heater, able to reach a maximum temperature of about 800°C, and a SiC-coated carbon susceptor. The susceptor can hold a single 6- or 4-inch wafer or three 2-inch wafers. The reactants reach the susceptor from the top via a close-coupled showerhead, which distributes the gas flow homogeneously over the surface of the wafer. From there the gases are vertically removed from the reactor through the side of the susceptor via a quartz liner system. The total pressure in the reactor can be varied from 30 torr to atmospheric pressure.

The precursor sources for the MOCVD growth are Trimethylgallium (TMGa) and Tertiarybutylarsine (TBAs). N-type doping is done with Silane, whereas P-type doping is achieved with Diethylzinc (DEZn). The carrier gas is hydrogen.

The GaAs analysed in this report is grown on bulk Ge 001 wafers. Ge has almost exactly the same lattice constant as GaAs. Only a 0.1% mismatch exists, which strongly facilitates epitaxial growth of GaAs on Ge. But the fact that GaAs is a polar material, which is grown on a non-polar substrate, makes GaAs epitaxy on Ge difficult, because anti phase domains (APDs) tend to develop with Ga-Ga and As-As bonds at the interfaces between the different domains [Kroemer 87]. Therefore, all the recipes for anti APD-free growth are based on the formation of double steps on miscut Ge substrates, and the formation of an As nucleation layer, which eliminates the problem of APDs [Pukite 87, Strite 89]. On non-miscut Ge wafers, GaAs growth without APDs is in general not possible.

For the following PL analysis we have deposited several tens of GaAs layers with thickness varying between 140 nm and 1 µm, different doping levels and on miscut as well as on non-miscut Ge 001 wafers. All layers were deposited at a temperature of 650°C.

## 4. Theoretical aspects for PL of GaAs on Ge

### 4.1. GaAs-Ge heterojunction

As we are looking into very thin GaAs films deposited on Ge substrates, it is of utmost importance to take a closer look at the effects of the formation of the Ge-GaAs heterojunction. With the PC1D software from the University of New South Wales [UNSW 05], we have calculated the band alignment, charge density, carrier density and electric field at the interface between the Ge and the GaAs. For the particular results displayed in figure 3, n-type doping of both layers of the order of $10^{16}$ states per $cm^3$ was assumed. Calculations were made for semiconductors at 100 K. Variations in temperature and doping concentration do not change the results remarkably. Using n- or p-type substrate has a strong effect on band alignment and charge type, of course, but the major conclusions from these calculations are not altered.

One can observe from the calculations that, due to the Fermi level mismatch between the two materials, the majority charge carrier from the GaAs will diffuse into the Ge, whereas the other charge carrier will diffuse from the Ge to the GaAs, no matter the type of the Ge. This leads to the creation of an electric field of the order of a Volt per µm in the vicinity of the heterojunction. If the GaAs is n-type, the electric field is positive and electrons will be repelled from the interface, whereas holes will be attracted by the interface. If the GaAs is p-type, the electric field is negative and electrons will be attracted by the interface, whereas holes will be repelled from the interface. The lateral extension of the electric field in the GaAs is approximately 200 nm. This therefore leads to an approximately 200 nm thick region close to the interface, where PL emission is very ineffective because of carrier separation due to the electric field.

## 4.2. Photon absorption and carrier diffusion in GaAs

When a photon with energy larger than the bandgap is absorbed in a semiconductor, an electron-hole pair is created. The excited electrons and holes relax to their respective band edges via phonon scattering in less than a picosecond [Elsaesser 92]. From here, the charge carriers can diffuse, before they finally recombine, thereby emitting a photon. The energy of this emitted photon is characteristic of the electronic state that the electron and hole were in. Non-radiative recombination via phonon release is possible as well, in which case no characteristic radiation is emitted [Pankove 71].

For 488 nm photons, the absorption coefficient α of GaAs is equal to $1.4 \cdot 10^5$ cm$^{-1}$ [Blakemore 82]. The ratio between transmitted light intensity $I_t$ and incoming light intensity $I_0$ through a film of thickness l can be expressed as:

$$\frac{I_t}{I_0} = \frac{1}{\exp(\alpha l)}. \tag{1}$$

From this we can calculate that about 85% of all incoming 488 nm photons are absorbed in a 140 nm thick GaAs film. For a 300 nm film the percentage is equal to 98.5 %. We can therefore state that pretty much all of the photons are absorbed in the first 300 nm of the material. Even for films thinner than 140 nm the absorbed proportion of photons is still reasonably high.

Once absorbed, the electrons and holes can diffuse freely for a timeframe equal to the recombination time. In GaAs of reasonably high quality the recombination time $\tau_r$ is of the order of 10 nsec [Lush 95], whereas the diffusion coefficient D is of the order of 20 cm$^2$/sec [Harmon 95]. The average diffusion length $l_d$ can then be calculated according to

$$l_d = \sqrt{\tau_r D}, \tag{2}$$

and is of the order of several µm in reasonably good quality GaAs. Therefore, in thin films, most of the charge carriers will easily reach the interface between the GaAs and Ge before recombining. According to the results of the previous section, at the interface, one type of charge carrier will be reflected by the electric field, whereas the other type of charge carrier will be attracted by the electric field and lost into the substrate.

# 5. PL from GaAs without APDs

## 5.1. General structure of PL spectra

The spectra acquired with our thin GaAs films on Ge substrates present two different structures. First of all, a narrow band-to-band (B2B) structure can be identified around the bandgap energy (~1.5 eV). These photons arise from recombination of free carriers, excitons or carriers trapped in shallow doping states [Pankove 71].
In addition to this peak, a very broad structure within the bandgap of the GaAs can be identified, at energies around 1.1 eV, which will be called the inner-bandgap (IB) structure in the following. These photons arise from the recombination of charge carriers, which are trapped in deep trapping states. These trapping states are usually charged structures within the GaAs composed of defects, impurity atoms or complexes of the latter [Reddy 96].
Figure 4 shows the spectra acquired at 77 K for three different GaAs films deposited on Ge, with thickness respectively equal to 140, 300 and 600 nm. The two different structures can be clearly identified.

### 5.1.1. Surface recombination

In order to learn more about the effect of surface recombination, we have grown a 1 μm thick GaAs film and deposited on top of the latter a 20 nm AlAs film. AlAs has a bandgap of 2.17 eV, larger than the bandgap of GaAs. Therefore, charge carriers in the GaAs are reflected at the GaAs-AlAs interface and thereby prevented from reaching the top surface, which is heavily oxidized and acts as a non-radiative recombination centre. The spectrum from such a film is shown in figure 5 along with the spectrum from a GaAs film that does not possess an AlAs top layer. Whereas the shape of the PL spectrum does not change considerably, small differences being due to the difference in film thickness, the intensity of the light emitted from the AlAs topped film is about a factor 40 higher than the intensity of the PL from the film without the AlAs. This shows that non-radiative surface recombination is indeed very strong in GaAs, which stresses the need for a proper GaAs surface passivation technique. Therefore, whenever high intensity is needed, for example for high resolution measurements, it is best to grow a thin AlAs layer on the top surface in order to reduce the surface recombination. In the following we will always indicate when such an AlAs film was grown in order to increase PL intensity.

## 5.2. Band-to-band (B2B) structure

The B2B structure depends strongly on the physical characteristics of the layers, like thickness and quality, as well as the doping level. For undoped films thinner than 200 nm, the B2B peaks change drastically from the structure of undoped films thicker than 200 nm. On the other hand, the structure of thicker films depends strongly on the doping level. We therefore classified the B2B structures into three different classes: thick undoped films, thick doped films and thin films.

### *5.2.1. Undoped GaAs on Ge*

Figure 6 shows the B2B structure of a 600 nm thick GaAs film grown on a Ge substrate measured at a temperature of 77K and with a spectrometer resolution of 0.4 meV.
The measured B2B structure consists of two peaks, which are separated by an energy of approximately 5 meV. Every one of theses two peaks consists itself of two components, one symmetrical, which is the contribution of excitonic recombination, and one asymetrical component, which corresponds to the free-carrier recombination. The splitting of the B2B peak into two separate peaks is due to the valence band splitting into a light hole and heavy hole component caused by stresses in the GaAs layer, as illustrated in figure 7 [Kuo 85].
Because of the 0.1 % difference in lattice constant between GaAs (5.6533 Å) and Ge (5.658 Å), the epitaxial growth of GaAs on Ge causes the GaAs to be stressed. From reciprocal space maps of films thinner than 1 µm we determined that the GaAs is fully stressed and adapts the lattice constant of the Ge in the in-plane direction, whereas it is compressed in the perpendicular direction, where it adapts a lattice constant of 5.6504 Å.
From this we can calculate the normal and parallel strain in the GaAs, which are respectively given by $\varepsilon_n = -5.8 \cdot 10^{-4}$ and $\varepsilon_p = 8.3 \cdot 10^{-4}$. We can deduce the biaxial tensile stress in the GaAs:

$$\sigma = C_{11}\varepsilon_p + C_{12}(\varepsilon_n + \varepsilon_p), \qquad (3)$$

where $C_{ij}$ are the elastic stiffness coefficients, whose numerical values for GaAs are respectively equal to $C_{11} = 119$ GPa and $C_{12} = 53.4$ GPa [Stolz 87]. This leads to a biaxial tensile stress in the GaAs equal to 112 MPa.
The valence band shifts due to the stress in the layer can then be calculated according to [Pikus 59, Pikus 60]:

$$\Delta E_{hh} = \left[ -2a \frac{C_{11} - C_{12}}{C_{11}} - b \frac{C_{11} + 2C_{12}}{C_{11}} \right] \varepsilon_n \qquad (4)$$

$$\Delta E_{lh} = \left[ -2a \frac{C_{11} - C_{12}}{C_{11}} + b \frac{C_{11} + 2C_{12}}{C_{11}} \right] \varepsilon_n, \qquad (5)$$

where $\Delta E_{hh}$ and $\Delta E_{lh}$ is the difference in bandgap energy with respect to the unstrained case for respectively the heavy hole and the light hole band, a is the hydrostatic and b is the shear deformation potential. The numerical values of the latter parameters for GaAs are [Stolz 87] a = -9.8 eV and b = -2.0 eV. Using these numerical values as well as the experimentally determined strain, we deduce differences in bandgap energy of respectively $\Delta E_{hh} = -8.5$ meV and $\Delta E_{lh} = -4.1$ meV. Knowing that the bandgap of unstrained GaAs is equal to 1.5075 eV at 77K, we deduce a bandgap energy of 1.499 eV for the heavy hole band and a bandgap energy of 1.5036 meV for the light hole band in the strained case. These values correspond exactly to the peak positions of the free carrier recombination peaks observed experimentally.
For the fit to the experimental data of figure 6 we used standard peak shapes. The shape of the excitonic recombination peak is described by the combination of a gaussian peak with the Boltzmann distribution function [Kudrawiec 03]:

$$I_x(\hbar\omega) = A_x \cdot \exp\left[ -\frac{(\hbar\omega - E_x)^2}{2\sigma_x^2} \right] \cdot \exp\left[ -\frac{(\hbar\omega - E_x)}{k_B T} \right], \qquad (6)$$

where $A_x$, $E_x$ and $\sigma_x$ are the amplitude, energy and broadening factor of the excitonic peak. The shape of the free carrier recombination peak is given by a broadened step-like density of states multiplied with the Sommerfeld factor and the Boltzmann distribution function [Kudrawiec 03]:

$$I_c(\hbar\omega) = A_c \cdot \frac{1}{1+\exp\left[-\frac{\hbar\omega - E_c}{\sigma_c}\right]} \cdot \frac{2}{1+\exp\left[-2\pi\sqrt{\frac{R^*}{\hbar\omega - E_c}}\right]} \cdot \exp\left[-\frac{(\hbar\omega - E_c)}{k_B T}\right], \quad (7)$$

where $A_c$, $E_c$ and $\sigma_c$ are the amplitude, energy and broadening factor of the free-carrier peak and $R^*$ is the exciton binding energy, which was taken equal to 2.5 meV for GaAs [Stolz 87]. All other parameters were used as fitting parameters for the fit of figure 6.

Measurements with temperature varying from 90K to 2.5K were made in order to gain some more insight into the behaviour of the different peaks. The measured film is a 1 μm thick undoped GaAs layer covered with a 20 nm AlAs top layer. For these measurements the resolution of the spectrometer was equal to 0.25 meV. Figure 8 shows the variation of the B2B response of the 1 μm film as the temperature is decreased from 90K to 2.5K. The different spectra are offset on the vertical axis in order to make the graph clearer. At 77 K the B2B structure is similar to the structure in figure 6, with two peaks, due to the heavy hole and the light hole band, each peak consisting of a symmetric excitonic recombination contribution and an asymmetric free-carrier recombination contribution. As the temperature is lowered, the bandgap of the GaAs increases, causing the peaks to shift to higher energies. In addition, the heavy hole peak intensity is reduced, whereas the light hole peak intensity increases. This arises because at lower temperatures the charge carriers condense at lower energies. Similarly, with decreasing temperature, the contribution due to the free-carrier recombination decreases. At these low temperatures all charge carriers are bound together into excitons at lower energies, causing the light hole peak to be purely gaussian at 20K. Below 20 K some more peaks start to appear, which correspond to excitons bound to impurity atoms. At least three of these impurity-bound exciton peaks can be observed at 2.5K. A somewhat higher resolution of the spectrometer, of the order of 0.1 meV or less, and lower temperatures would be required for the determination of the exact number of lines in the spectrum. Nevertheless the most dominant bound exciton peak can be identified approximately 3 meV below the free exciton peak and corresponds to the carbon-bound exciton recombination [Hamilton 96].

The energy resolutions of the different peaks in the low temperature spectra, which have line widths well below one meV, show the very good structural quality of the GaAs grown on the Ge substrates. The peak widths are comparable to the ones observed in very high quality epitaxial GaAs grown on bulk GaAs substrates [Gilleo 68] and much better than for similar GaAs films grown on Ge substrates, which show low temperature peak widths of the order of 5 meV [Knuuttila 05, Modak 98, Hudait 98, Hudait 01].

### 5.2.2. *Doped GaAs on Ge*

As the doping level is increased the PL spectra change according to the change of the density of states of the charge carriers. We have doped our layers with Silane, resulting in an n-type doping that is approximately proportional to the ratio of the Silane molar flow to the TMG molar flow through the reactor [Stringfellow 99].

Figure 9 shows the B2B structures of four different GaAs layers with a thickness of 1 μm and topped with a 20 nm AlAs film. The ratio of Silane molar flow to TMG molar flow during growth of the four films was respectively 0, 1.1 10$^{-4}$, 4.3 10$^{-3}$ and 1.4 10$^{-2}$. In the figure, the intensity of the spectrum of the film with the highest Silane flow was magnified by a factor 500, because of the very low PL intensity of this film, as compared to the other films.

One can clearly identify that the PL peaks broaden as the Silane flow is increased. An empirical formula relates the full width at half maximum (FWHM) resolution of the peak ΔE to the n-type doping concentration n in the film [De-Sheng 82]:

$$\Delta E = 3.84 \cdot 10^{-14} n^{\frac{2}{3}}, \tag{8}$$

where ΔE is expressed in eV and n is expressed in cm$^{-3}$. Using the FWHM from the spectra equal to 13, 47 and 113 meV, we obtain doping levels of 2 10$^{17}$, 1.4 10$^{18}$ and 5 10$^{18}$ cm$^{-3}$ for the films with molar ratios of respectively 1.1 10$^{-4}$, 4.3 10$^{-3}$ and 1.4 10$^{-2}$. These values correspond reasonably well to the values obtained for GaAs deposited on GaAs substrates as can be seen in figure 10. The blue data points represent carrier concentration data determined from Hall measurements for a series of doped GaAs layers deposited on GaAs substrates. The red data points represent the carrier concentrations of GaAs films deposited on Ge substrates, as determined from the FWHM of the B2B transitions in the PL spectra. Whereas the uncertainty on the carrier concentration value for this kind of measurement is rather large, it still gives an impression about the doping concentration in the layer.

Similarly, it can be identified that the peak position of the B2B structure gradually moves to higher energies as the doping level is increased. This well-known feature is the Burstein-Moss effect arising from the filling of the conduction band as the doping level is increased [Burstein 54, Moss 54].

The highest doping level achievable with Si doping is approximately 5 10$^{18}$ cm$^{-3}$. At this doping level the Si impurity atoms start to occupy the As positions in the GaAs lattice, which act as acceptors and therefore compensate the effect of the donor atoms [Kressel 69]. We expect that the highest doped film, grown with a molar ratio of 1.4 10$^{-2}$, is in this compensation regime caused by the amphoteric character of the Si atoms in the GaAs lattice. This suspicion is confirmed by the very low intensity of the B2B peak, which is a factor 1000 lower than the film with a slightly lower doping concentration. In this film the charge carriers must therefore recombine through some other mechanism, probably related to Si impurities occupying the As positions in the GaAs lattice. We will come back at this in the section discussing the effect of the doping level on the structure of the inner bandgap peaks.

### 5.2.3. Thin GaAs on Ge

For undoped GaAs films grown on Ge substrates thinner than 200 nm, the spectrum changes from the shape of a typical undoped spectrum to the shape of a doped spectrum with a FWHM of the B2B peak larger than 10 meV. Figure 11 shows the B2B spectrum of a 140 nm thick GaAs film deposited on Ge with a 20 nm thick AlAs top layer. The FWHM of the B2B peak is equal to 56 meV, which corresponds to the PL signal of a very highly doped layer. Using equation (8) the determined doping level of the film is equal to 2 10$^{18}$ cm$^{-3}$. The intrinsic doping of this unintentionally doped film probably arises from heavy in-diffusion of Ge from the substrate into the epitaxial GaAs film. SIMS measurements on layers deposited at similar growth conditions than the present films have shown that indeed the Ge diffuses as far as 250 nm into the GaAs [Hudait 98]. These measurements are therefore consistent with the present results.

## 5.3. Inner-bandgap (IB) structure

The broad IB structure is in general composed of two gaussian peaks centred around 1.04 and 1.17 eV and with FWHM of about 160 meV. Figure 12 shows the IB structures of the three undoped GaAs films with thickness equal to 140, 300 and 600 nm, whose full spectra are shown in figure 4. For every structure a fit to the experimental data is shown as well as the individual gaussian contributions to the IB peak.

PL studies of Si-doped GaAs [Kressel 67, Souza 90, Okano 89, Visser 91, Hudait 98] have shown that deep trapping states similar to the ones observed in our films are formed when the GaAs is very heavily doped at levels in excess of $5 \cdot 10^{18}$ cm$^{-3}$. At this high doping level the GaAs reaches the compensation regime, where additional Si atoms, which usually are donors and replace Ga atoms in the GaAs lattice, start to occupy the position of the As atoms, which leads to a certain amount of acceptors in the layer. When this compensation regime is reached, the doping density stabilises, because of the mutual elimination of the effects of Si atoms sitting at Ga positions Si$_{Ga}$ and Si occupying the As positions Si$_{As}$ in the GaAs lattice. The exact nature of the deep trapping states observed is still controversial, but in general the consensus is that a 1.05 eV peak is related to the Si$_{Ga}$-Si$_{As}$ complex [Visser 90, Hudait 98], whereas a 1.2 eV peak is related to the Si$_{Ga}$-V$_{Ga}$ complex [Souza 90, Visser 90, Hudait 98], where V$_{Ga}$ is a Ga-vacancy site.

### 5.3.1. Variations with film thickness

As can be clearly observed on figure 12, the intensity of the IB peaks increases as the film thickness is decreased. The integrated PL intensity of the IB structure is shown in figure 13 as a function of the thickness t of the three films. Along with the experimental data, three general dependencies are shown, corresponding to the $1/t$, $1/t^2$ and the $1/t^3$ dependencies. It can be observed that the general dependency of the IB PL intensity is somewhere between $1/t$ and $1/t^2$. More data points need to be collected in order to be able to get a clearer view of the exact dependency of the IB-PL intensity on the film thickness.

Because of this strong dependency on the film thickness, it can be stated that the IB structure originates mainly at the GaAs/Ge interface, because of very heavy in-diffusion of Ge into the GaAs. As the film thickness increases, less charge carriers are able to reach the GaAs/Ge interface by diffusion before recombination and the intensity of the IB structure decreases accordingly. Because the dependency is not directly correlated to the inverse of the film thickness, but rather to the thickness squared, diffusion is probably not the only factor limiting the transport of the charge carriers. Probably, the electric field created by the heterojunction between the GaAs and the Ge also plays a role and prevents the charge carriers in the thicker films to reach the interface, which gives an additional reduction of the IB intensity for thicker films. Clearly, in order to get a better view of the mechanisms involved, more data points on the variation of the integrated PL intensity of the IB peak with film thickness are needed.

### 5.3.2. Variations with doping level

Figure 14 shows the IB structures of four 1 µm thick GaAs films topped with 20 nm of AlAs, doped with Silane flows of respectively 0, 12, 450 and 1430 nmol/min. Whereas for the three lowest doping levels the IB structure does not change much, there is a clear influence of the doping on the IB structure for the film with the doping level for which the compensation regime is reached.

For the three lowest doping levels, the Si doping atoms predominantly occupy the $Si_{Ga}$ positions in the lattice. Because of a lack of $V_{Ga}$ and $Si_{As}$ impurity states in our high quality material, no complexes can form that would create deep level trapping states. Therefore, the increase in Si doping, for the low doping levels, does not vary the IB emission. The emission that can be identified on figure 14, for the low doping levels $n < 5 \; 10^{18}$ cm$^{-3}$, is probably solely due to impurities at the Ge/GaAs interface. It is known that for the growth temperatures at which our films were grown, around 650°C, Ge diffuses strongly into the GaAs with Ge impurity levels in excess of $5 \; 10^{18}$ cm$^{-3}$ for the regions very close to the Ge substrate [Hudait 98]. At the interface $Ge_{Ga}$-$V_{Ga}$ and $Ge_{Ga}$-$Ge_{As}$ complexes can form, which probably behave very similarly to the corresponding Si complexes that have been more widely studied. The only difference seems to be that the energy levels of the Ge-based complexes at 1.04 and 1.17 eV seem to be slightly lower as compared to the Si-based complexes at 1.05 and 1.19 eV.

As the Si impurity level is increased above the compensation region limit, the IB structure changes. Whereas the peaks at 1.04 and 1.17 eV are slightly reduced in intensity and shifted towards slightly higher energy levels, a third peak can be identified at an energy of about 1.27 eV. This indicates that the light does in this case not come from the interface anymore, but rather from the Si-based complexes that now formed in the bulk of the GaAs film. Diffusion of charge carriers to the GaAs/Ge interface is prevented by the charged, deep level trapping states $Si_{Ga}$-$Si_{As}$ and $Si_{Ga}$-$V_{Ga}$, which act as strong recombination centres, reducing the diffusion constant. At the same time these Si-based complexes emit their characteristic light, which has slightly higher energies than for the Ge-based complexes. The additional peak at around 1.27 eV seems to arise from a $V_{As}$-$As_i$ complex, where $As_i$ is an As interstitial atom [Reddy 96].

## 6. PL from GaAs with APDs

All previously analysed films were deposited on 6 degree miscut Ge wafers. We also deposited GaAs films on 0 degree miscut Ge substrates. It is known that on these non-miscut wafers, the polar GaAs shows APDs [Kroemer 87]. Figure 15 shows a Nomarski microscope image of a 1 μm thick GaAs film grown on a non-miscut Ge substrate after defect etching, which visualizes the APDs in the material.

The boundaries of these APDs consist of Ga-Ga and As-As bonds, which are electrically very active centres in GaAs and act as non-radiative recombination centres.

Figure 16 shows the PL spectra of three 1 μm thick GaAs films deposited on non-miscut wafers, which contain such APDs. The three films are Si-doped with ratio of Silane molar flow to TMG molar flow during growth of respectively 0, $1.1 \; 10^{-4}$ and $4.3 \; 10^{-3}$. For comparison, the spectrum of a 1 μm thick undoped film grown on 6° miscut Ge, therefore without APDs, is shown in the figure as well. In order to show the full variation of the PL intensity for the different samples, the spectra are shown on a logarithmic scale.

It is clear that the effect of the APDs is to suppress most of the light emission from the GaAs films. The intensity from an undoped film without APDs is four orders of magnitude larger than for a film with APDs. As a matter of fact the Ga-Ga and As-As are electrically very active and act as strong non-radiative recombination centres. It is remarkable to observe though that with increasing Si doping level, the intensity of the PL signal increases considerably again. When the Si doping level is approximately $10^{18}$ cm$^{-3}$, the PL intensity of a film with APDs is only a factor 10 lower than for the same film without APDs. The Si impurities seem to condense at the anti phase boundaries and annihilate the effect of the Ga-Ga and As-As bonds. Another indication towards the fact that the Si impurities accumulate at these anti-phase boundaries is the fact that the IB peak at 1.17 eV is more strongly suppressed than the IB peak at 1.04 eV. As the Si and Ge impurities condense at the anti phase boundaries, the $Ge_{Ga}$-$Ge_{As}$ complex can still be formed and its density might even increase as the Ge atoms are gathered in clusters near the anti phase boundaries. The $Ge_{Ga}$-$V_{Ga}$ complex, on the other hand, has less probability to form in these films, as the Ga vacancies are not influenced by the APDs and are distributed randomly inside the bulk, whereas the Ge still accumulates at the boundaries between the different phase domains. As a consequence the ratio between the intensity of the peak at 1.17 eV to the intensity of the peak at 1.04 eV decreases in films with APDs. Figure 17 shows graphically the ratio between the intensity of the peak at 1.17 eV and the intensity of the 1.04 eV film, for 20 different undoped films deposited on Ge. Whereas the physical characteristics like thickness and structural quality of all films are completely different, it can be identified that the ratio is dependent on the existence of APDs in the films. For films without APDs the intensity ratio is in general larger than 1.4, whereas it is smaller than 1.4 in films, which show the presence of APDs during defect etching studies.

## 7. Conclusions

We presented an extensive, low-temperature photoluminescence (LTPL) study of thin GaAs films epitaxially grown by metal organic vapor phase deposition on Ge substrates. GaAs thin films were grown without anti phase domains on 6° miscut Ge substrates, as well as with anti phase domains on 0° miscut Ge. Films with layer thickness varying between 140 nm and 1 μm and different doping concentration levels have been analysed.

A study of the GaAs/Ge heterojunction was made, which reveals the existence of an inherent electric field at the interface that influences the kinetics of the charge carriers in the vicinity of the interface. Whereas the GaAs majority charge carriers are repelled by the interface, the minority charge carriers are attracted by the interface and drained into the Ge substrate. The top surface of the GaAs films, on the other hand, acts as a strong non-radiative recombination centre. Covering the GaAs films with 20 nm of AlAs increases the PL intensity by a factor 30.

All PL spectra of thin GaAs deposited on Ge show two different structures: One band-to-band (B2B) structure at an energy of approximately 1.5 eV and a broad inner-bandgap (IB) structure at an energy of about 1.1 eV.

The B2B structure of thin undoped GaAs films on Ge consists of two peaks, red-shifted by respectively 8.5 and 4 meV from the bulk GaAs case. This energy shift and double peak structure is due to strain in the GaAs layer grown on Ge, which causes a valence band splitting into a light- and a heavy-hole band. At 77K every peak in the B2B structure consists itself of two contributions corresponding respectively to the free-carrier and the excitonic recombination. At temperatures below 10 K several bound-exciton peaks appear, the dominant-one being the C-bound exciton peak at energy approximately 3 meV below the free-exciton peak. At 2.5 K the FWHM resolutions of the different excitonic peaks are well below 1 meV, showing the very good structural quality of the GaAs grown on the Ge substrates, comparable to very high quality GaAs grown on GaAs substrates.

The B2B structure of Si-doped GaAs on Ge shows the usual dependence of the FWHM of the band-to-donor peak on the doping level. In addition, for high doping levels, an increase of the peak energy is observed with increasing doping level, due to the Burstein-Moss shift. For Si-doping levels larger than $5 \cdot 10^{18}$ cm$^{-3}$, the GaAs enters the self-compensation regime. In this regime Si atoms take the As positions in the GaAs lattice and electrically active deep-level states consisting of complexes of Si impurities are formed. Most recombination is then mediated vie these very active recombination centres and the intensity of the B2B structure is reduced by three orders of magnitude.

For undoped GaAs films deposited on Ge with thickness smaller than 300 nm, the B2B structures correspond to the ones of heavily n-type doped films. This n-type doping probably arises from in-diffusion of Ge into the GaAs, which was already previously observed via SIMS studies. For a 140 nm thick GaAs film, a doping level of $2 \cdot 10^{18}$ cm$^{-3}$ was determined from the peak width, which is in agreement with the results of the previous SIMS studies.

The IB structure observed in all films consists of two broad peaks with peak energies at 1.04 eV and 1.17 eV. These are created by deep-level trapping states at the interface between the Ge and the GaAs. Due to the strong Ge in-diffusion into the GaAs, complexes of Ge impurities form at the interface. The peak at 1.04 eV seems to be related to the $Ge_{Ga}$-$Ge_{As}$ complex, whereas the peak at 1.17 eV seems to arise from recombination through the $Ge_{Ga}$-$V_{Ga}$ complex. The intensity of the IB structure decreases as the thickness of the GaAs film increases, because less charge carriers are able to diffuse to the interface. The IB structure does not vary with Si doping level, as long as the doping level is kept below $5 \cdot 10^{18}$ cm$^{-3}$. For higher doping levels the compensation regime is attained and the IB structure of self-compensated Si-doped GaAs becomes visible, with broad peaks at 1.05, 1.19 and 1.27 eV. Due to the large recombination cross-sections of these impurity complexes, the charge carriers are not able to diffuse to the interface and the peaks caused by the Ge-based complexes disappear.

The 77 K PL intensity of undoped GaAs films grown on 0° miscut Ge substrates, therefore containing anti phase domains (APDs), is reduced by 4 orders of magnitude as compared to similar films grown on 6° miscut substrates without APDs. This strong decrease can be explained by the electrically very active Ga-Ga and As-As bonds present at the anti phase boundaries, which act as strong non-radiative recombination centres. With increasing Si doping the intensity of the PL light increases again. For a film with Si doping of approximately $10^{18}$ cm$^{-3}$, the PL intensity is only a factor 10 lower as compared to a similar film without the APDs. We tentatively explain this phenomenon by Si-impurity clustering at the anti phase boundaries, which neutralizes the effect of the Ga-Ga and As-As bonds. This hypothesis is confirmed by the fact that in films including APDs, the intensity of the IB peak at 1.17 eV is stronger reduced than the intensity of the peak at 1.04 eV. Due to the Si impurity clustering at the anti phase boundaries the density of the $Si_{Ga}$-$Si_{As}$ complex increases, whereas the $Si_{Ga}$-$V_{Ga}$ complex has less probability to form. This is translated into a higher relative intensity of the 1.04 eV peak as compared to the 1.17 eV peak. The ratio of the intensity of the 1.17 eV peak to the intensity of the 1.04 eV peak is larger than 1.4 in all undoped films without APDs, whereas it is smaller than 1.4 in all undoped films containing APDs, this irrespective of the physical characteristics of the layers.

**Figures**

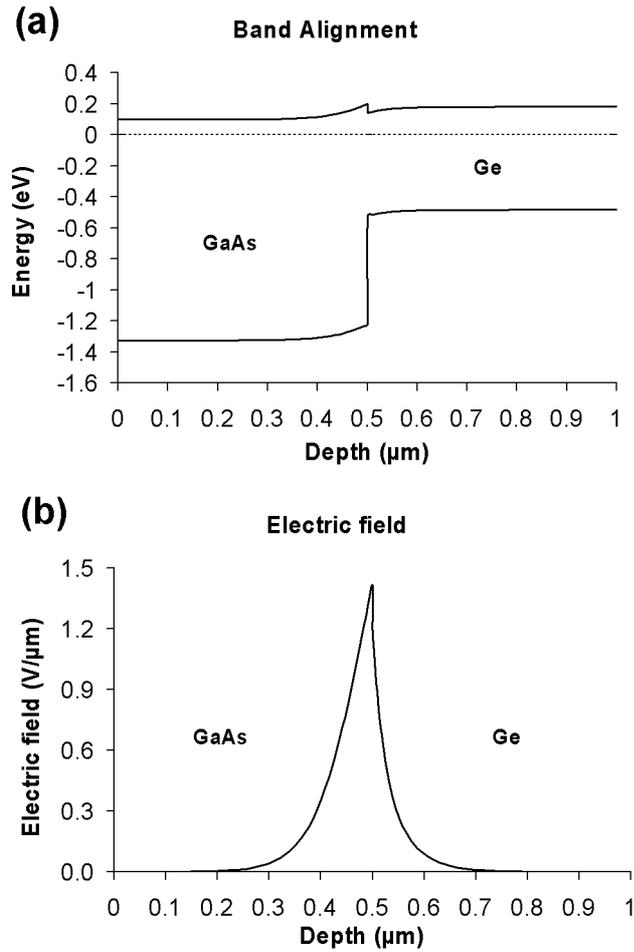

**Figure 1:** Calculation of band alignment (a) and electric field (b) for the heterojunction between n-type GaAs ($10^{16}$ atoms/cm$^3$) and n-type Ge ($10^{16}$ atoms/cm$^3$).

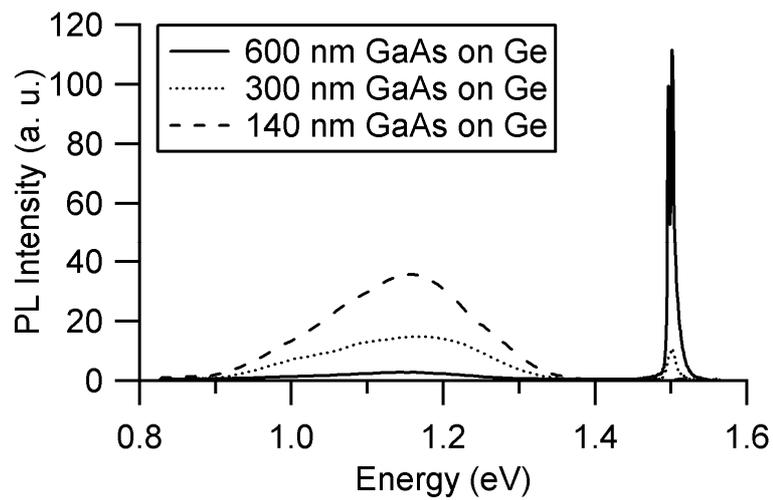

**Figure 2:** 77K PL spectra of three undoped GaAs films deposited on Ge with GaAs film thickness respectively equal to 140 nm (dashed line), 300 nm (dotted line) and 600 nm (solid line).

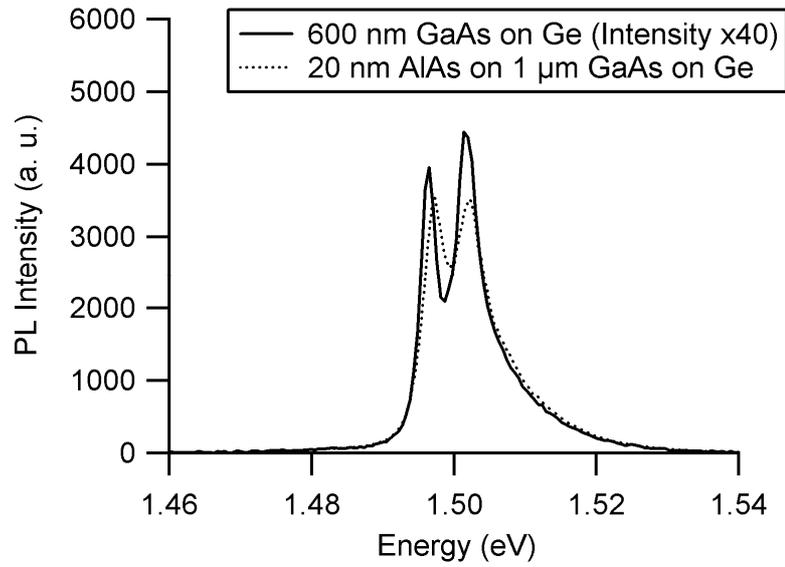

**Figure 3:** 77K B2B PL spectrum of a 1 μm thick GaAs film with a 20 nm AlAs top layer (dotted line) and a 600 nm GaAs film on Ge (solid line). The PL intensity of the 600 nm thick film was multiplied by a factor 40.

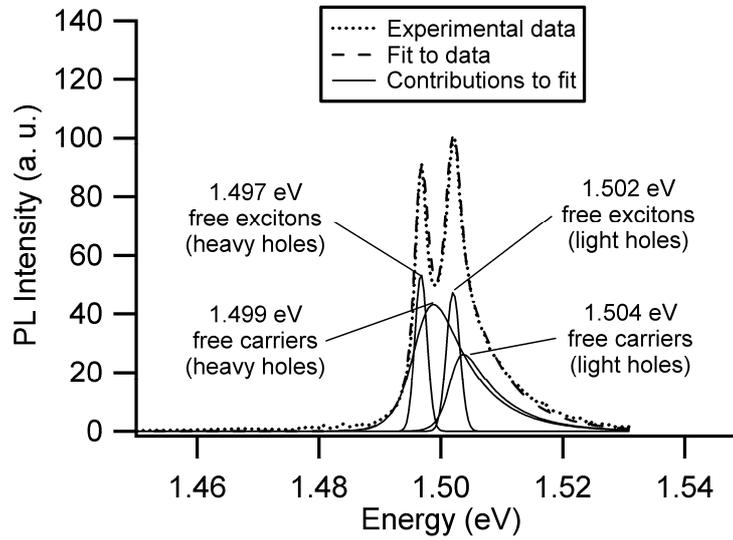

**Figure 4:** 77K B2B PL spectrum of a 600 nm thick GaAs film on Ge (dotted line). A fit to the experimental data is shown as well (dashed line). The solid lines show the different components of the fit.

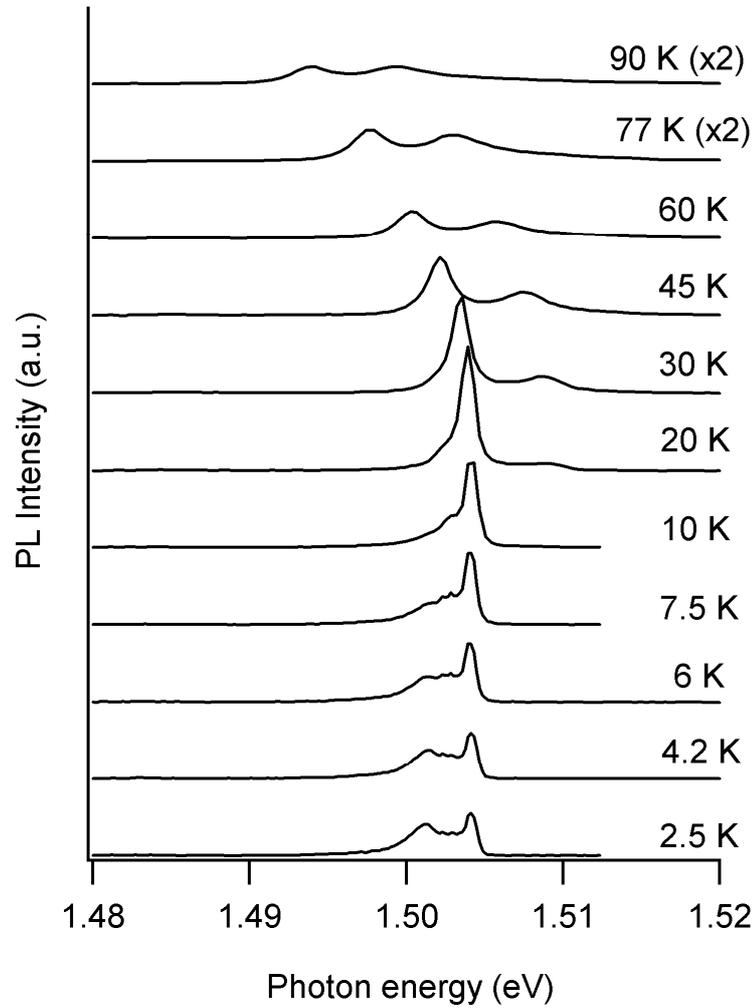

**Figure 5:** Variation of the B2B PL intensity of a 1 μm GaAs film covered with a 20 nm AlAs layer as a function of temperature. The spectra at different temperatures are offset on the vertical axis for clarity.

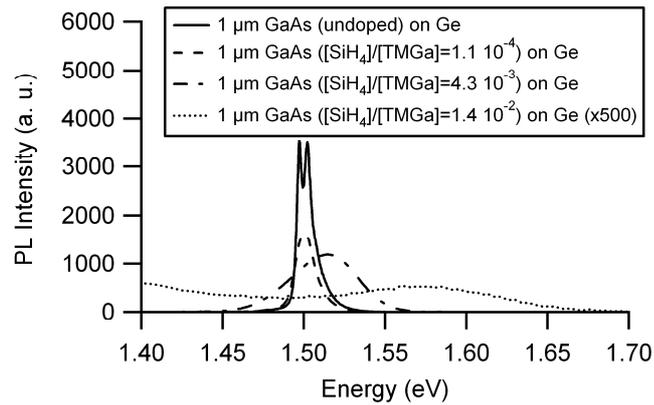

**Figure 6:** 77K B2B PL spectra of four 1 μm thick GaAs films topped with 20 nm of AlAs, doped with ratio of Silane molar flow to TMGa molar flow of respectively 0, $1.1 \cdot 10^{-4}$, $4.3 \cdot 10^{-3}$ and $1.4 \cdot 10^{-2}$. The intensity of the film with the highest Si doping was magnified by a factor 500 in order to improve visibility.

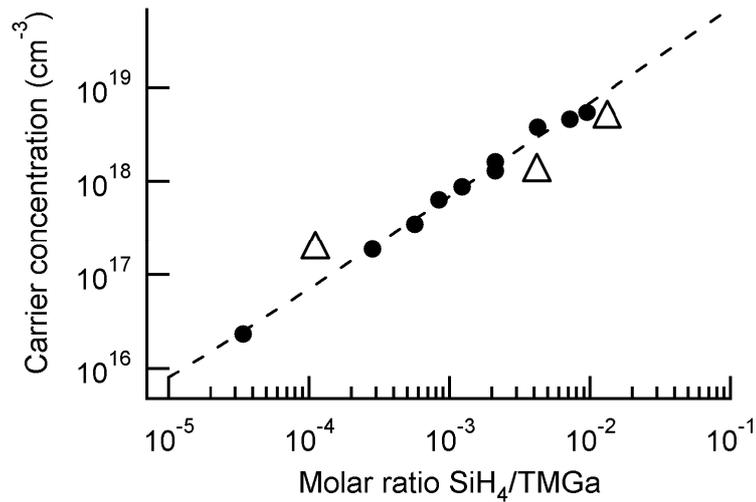

**Figure 7:** Carrier concentration versus ratio of Silane molar flow to TMGa molar flow. The circles represent calibration data from Hall measurements of GaAs films grown on GaAs substrates and the dashed line is a fit to that data. The triangles are the values determined from PL measurements of GaAs films grown on Ge substrates.

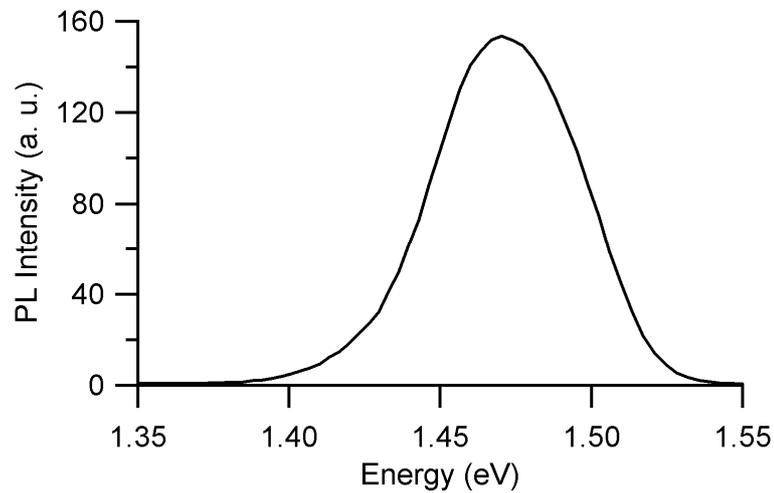

**Figure 8**: 77K B2B PL spectrum of a 140 nm thick undoped GaAs layer covered with 20 nm of AlAs.

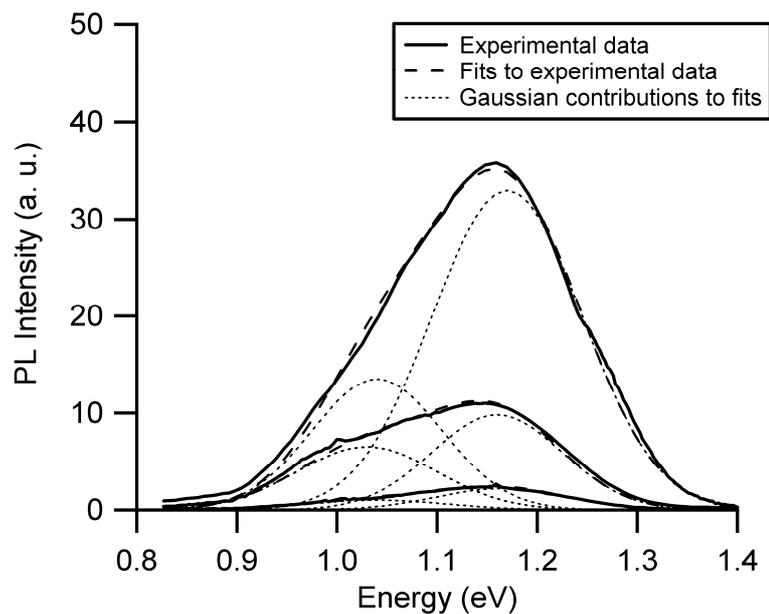

**Figure 9:** The solid lines represent the 77 K IB PL spectra of three GaAs films with different thickness on Ge substrates. The top solid curve is the IB spectrum of a 140 nm thick film, the middle curve corresponds to a

300 nm thick film and the bottom curve corresponds to a 600 nm thick GaAs film on Ge. For every structure a fit to the experimental data is shown (dashed curves) as well as the individual gaussian contributions to the fit with peak positions at 1.04 and 1.17 eV (dotted curves).

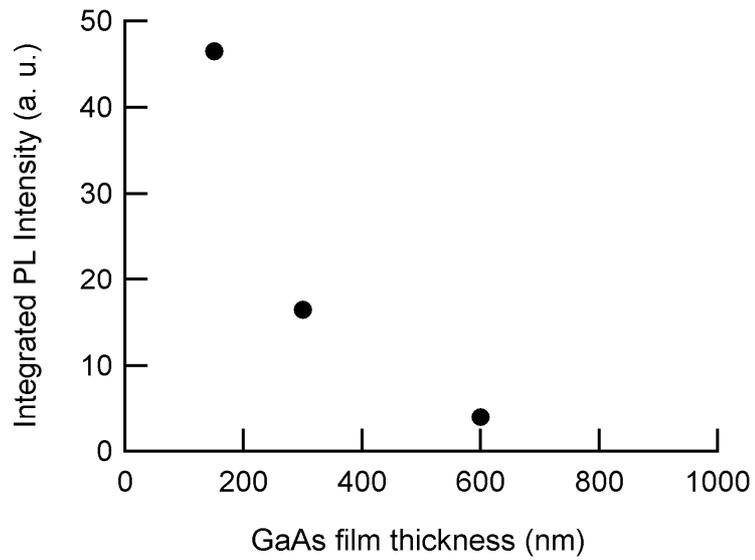

**Figure 10:** Variation of the integrated PL intensity of the IB structure as a function of the GaAs film thickness.

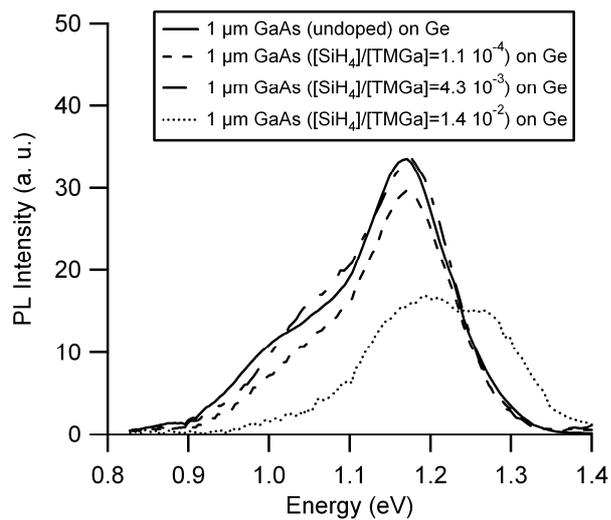

**Figure 11:** 77 K IB PL spectra of four 1 μm thick GaAs films topped with 20 nm of AlAs, Si-doped with Silane molar flow to TMGa molar flow of respectively 0, $1.1\ 10^{-4}$, $4.3\ 10^{-3}$ and $1.4\ 10^{-2}$.

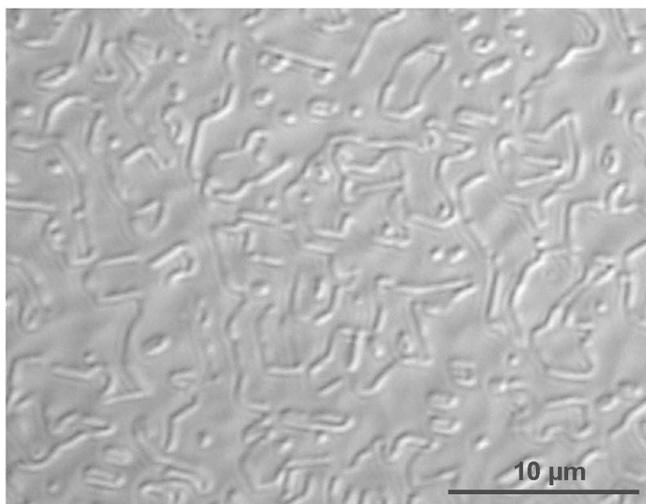

**Figure 12:** Nomarski microscope image of a 1 µm thick GaAs film grown on a 0° miscut Ge wafer after defect etching, revealing APDs in the material.

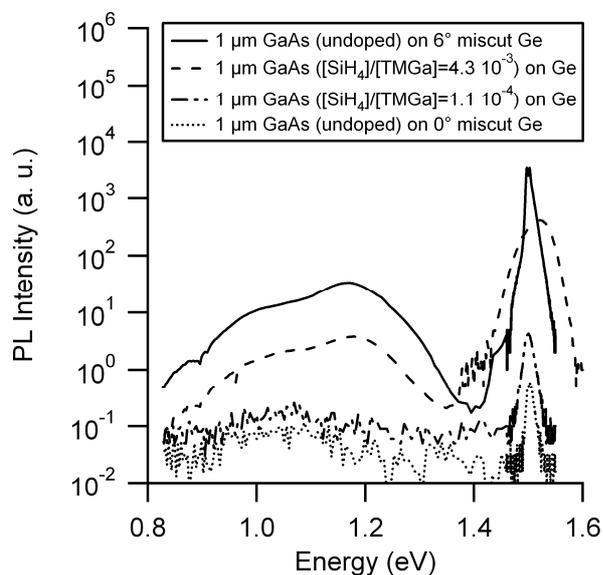

**Figure 13:** 77 K PL spectra of 1 µm thick GaAs layers with three different Si-doping levels deposited on 0° miscut Ge, as compared to an undoped film deposited on 6° miscut Ge. All films are covered with a 20 nm AlAs layer in order to prevent surface recombination.

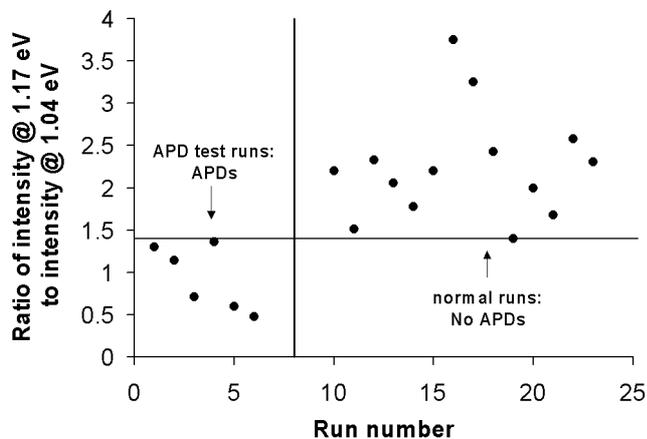

**Figure 14:** Ratio of the intensity of the 1.17 eV peak to the intensity of the 1.04 eV peak for 20 different GaAs films on Ge substrates. All films with APDs are shown on the left of the vertical line, whereas all films without APDs are plotted on the right of the vertical line.